\documentstyle[epsfig]{aipproc}
\newcommand{\epsfigure}[2]{\epsfig{file=#1,width=#2}}
\newcommand {\bi} {\bibitem}
\newcommand {\be} {\begin{equation}}
\newcommand {\ee} {\end{equation}}
\newcommand {\bea} {\begin{eqnarray} }
\newcommand {\eea} {\nonumber \end{eqnarray}}
\newcommand {\eps} {\epsilon}
 \newcommand {\si} {\sigma}

\newcommand {\ba} {\overline}
\newcommand {\lan} {\langle}
\newcommand {\ran} {\rangle}

\newcommand {\bc} {\begin{center}}
\newcommand {\ec} {\end{center}}
\newcommand {\bd}{\begin{displaymath}}
\newcommand {\ed}{\end{displaymath}}

\def \form#1 {eq. (\ref{#1}) }
\def \parziale#1#2  {{\partial {#1} \over \partial {#2}}}

\begin{document}
\title{Stochastic Stability}

\author{Giorgio Parisi}
\address{Dipartimento di Fisica, Sezione INFN and Unit\`a INFM,\\
Universit\`a di Roma ``La Sapienza'',
Piazzale Aldo Moro 2,
I-00185 Rome (Italy)}

\maketitle

\begin{abstract}
In this talk I will introduce the principle of stochastic stability and discussing its consequences 
both at equilibrium and  off-equilibrium.
\end{abstract}

\section*{Introduction}
In this talk I will underline the physical meaning of replica symmetry breaking 
\cite{MPV,PABOOK,CINQUE}. In this framework I will introduce the principle of Stochastic  Stability: 
I will present its far reaching consequences on the equilibrium properties of the system and on the 
off-equilibrium behaviour.

\section*{The coexistence of many phases}

Usually, if a system has different phases which are separated by a first order transition, just at 
the phase transition point a very interesting phenomenon is present: phase coexistence.  This 
usually happens if we tune one parameter.  This behaviour is summarized by the Gibbs rule which 
states that, in absence of symmetries, we have to tune $n$ parameters in order to have the 
coexistence of $n+1$ phases.

In the case of complex systems we have that the opposite situation is valid: the number of phases is 
very large for a generic choice of parameters.  It is usual to assume that all these states are 
globally very similar: states can be separated only by comparing one state with an other.  An 
example of this phenomenon would be a very long heteropolymer, e.g, a protein or RNA, which folds in 
many different structures.  However quite different foldings  may have a very similar 
density.  Of course you will discover that two proteins have folded in two different structures if 
we compare them.

In order to be precise we should consider a large but ({\sl finite}) system \cite{CINQUE}.  We want 
to decompose the phase state in valleys (phases, states) separated by barriers. If the free energy 
as function of the configuration space has many minima, the number of states will be very large.

Let us consider for definitiveness  a spin system with $N$ points
(spins are labeled by $i$, which in some cases will be a lattice point).
States (labeled by $\alpha$) are characterized by different local  magnetizations:
$ 
m_{\alpha}(i) =\lan \sigma(i) \ran_{\alpha}  \  ,
$
where $\lan \cdot \ran_{\alpha}$ is the expectation value
in the valley labeled by $\alpha$.
The average done with the Boltzmann distribution is denoted as $\lan \cdot \ran$ and it 
can be written as linear combinations of the averages inside the valleys. 
We have the relation:
\be
\lan \cdot \ran \approx \sum_{\alpha} w_{\alpha} \lan \cdot \ran_{\alpha}\ . \label{DECON}
\ee
We can also write  the relation
$
w_{\alpha} \propto  \exp(-\beta F_{\alpha})  
$,
where by definition $F_{\alpha}$ is the free energy of the valley labeled by $\alpha$.

In the rest of this talk I will call $J$ the control parameters of the systems.
The average over $J$  will be denoted by  a bar (e.g. $\ba{F}$).
I will consider here the case where a quenched disorder is present: 
the variables $J$ parametrize  the quenched disorder.

\section*{The overlap and its probabilities}

We have already remarked, states may be 
separated making a comparison among them. At this end it is convenient to consider their 
mutual overlap. 
Given two configurations, we define their overlap:
\be
q[\sigma,\tau]= \frac{1}{N} \sum_{i=1,N}\sigma(i)\tau(i) \ .
\ee
The overlap among the states is defined as
\be
q(\alpha,\gamma)=\frac{1}{N} \sum_{i=1,N}m_{\alpha}(i)m_{\gamma}(i)\approx
q[\sigma,\tau]\ ,
\ee
where $\sigma$ and $\tau$ are two generic configurations that belong to the states $\alpha$ 
and $\gamma$ respectively.

We define $P_{J}(q)$ as probability distribution of the overlap  $q$ at given $J$, 
i.e. the histogram of $q[\si,\tau]$, where $\si$ and $\tau$ are two equilibrium 
configurations. Using eq. (\ref{DECON}), one finds that

\be P_{J}(q)=\sum_{\alpha,\gamma}w_{\alpha}w_{\gamma}\delta(q-q_{\alpha,\gamma}) \ ,
\ee
where in a finite volume system the delta functions are smoothed.
If there is more than one state, $P_{J}(q)$ is not a single delta function:
\be
P_{J}(q)\ne \delta (q-q_{EA}) \ .
\ee
If this happens we say that the replica symmetry is broken: two identical replicas of the 
same system may state in a quite different state. However at zero magnetic field 
$P_{J}(q)$ is an even function, so that at low temperature it will contains two delta functions at 
$|q|\ne0$.
\begin{figure}[t] % fig 1
\begin{center}
\epsfigure{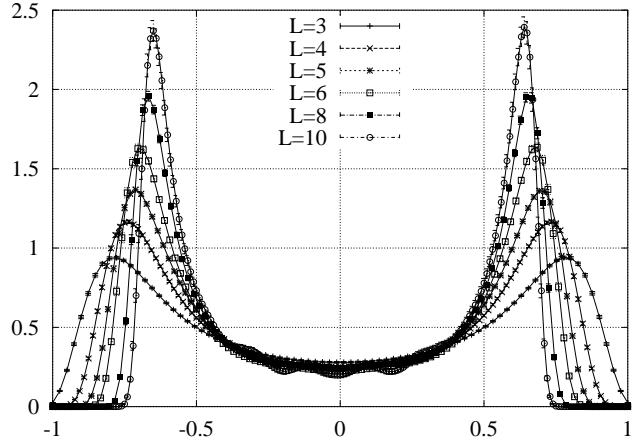}{3.5in}
\end{center}
\vspace{10pt}
\caption{The function $P(q)$ after average over many samples in four dimensions for system of size 
$L^{4}$, (L=3\ldots 10)  .}
\label{D4}
\end{figure}

There are many models where the function $P_{J}(q)$ is non trivial: 
a well known example is given by Ising spin glasses \cite{MPV,BY,FiHe}. In this case the Hamiltonian
is given by

\be H=-\sum_{i,k}J_{i,k}\sigma_{i}\sigma_{k} -\sum_{i}h_{i} \si_{i}\ ,
\ee
where $\si=\pm 1$ are the spins. The variables $J$ are random couplings (e.g. Gaussian or 
$\pm 1$) and the variables $h_{i}$ are the magnetic fields, which may be point dependent.

Let is consider two different models for spin glasses:

a) The Sherrington Kirkpatrick model (infinite range): all $N$ points are connected: 
$J_{i,k}=O(N^{-1/2})$.  Eventually $N$ goes to infinity.

b) Short range models: $i$ belongs to a $L^{D}$ lattice. The interaction is nearest neighbour 
(the variables $J$ are or zero or of order 1) and eventually 
 $L$  goes to infinity at fixed $D$ (e.g. $D=3$).

Analytic studies have been done in the case of the SK model, where one can prove 
rigorously that the function $P_{J}(q)$ is non-trivial. In the finite dimensional case a homologous
theorem has not been proved and in order to answer to the question if the function $P_{J}(q)$
is trivial we must resort to some numerical simulations \cite{romani-young} or to 
experiments.

In this case numerical simulations present ample evidence that in three and four dimensions the 
function $P_{J}(q)$ is non-trivial, i.e. at zero magnetic field it is not the sum of two delta 
functions: this can be seen   looking to the average of $P_{J}(q)$,
shown fig. (\ref{D4}) from \cite{YYY}. We will see later that the
the function $P_{J}(q)$ strongly depends on $J$.

\section*{Stochastic stability}
In the nutshell stochastic stability states that the system  we are considering behaves 
like a generic random system \cite{guerra,aizenman,PARI}.
Technically speaking in order to formulate stochastic stability we have to consider the 
statistical properties 
of the system with Hamiltonian given by the original Hamiltonian ($H$) plus a random 
perturbation ($H_{R}$):
\be
H(\eps)=H+\eps H_{R} \ .
\ee
Stochastic stability states that all the properties of the system are smooth functions
of $\eps$ around $\eps=0$, after the appropriate
averages
over the original  Hamiltonian and the random Hamiltonian.

Typical examples of random perturbations are (we can chose the value of  $r$ in 
an arbitrary way):

\begin{equation}
H_{R}^{(r)}=N^{(r-1)/2}\sum_{i_{1}\ldots i_{r}} R({i_{1}\ldots i_{r}}) \si(i_{1}) \ldots 
\si(i_{r}) \ ,
\end{equation}
where for simplicity we can restrict ourselves to the case where the variables $R$ are 
random uncorrelated Gaussian variables.

\begin{figure}[t] 
\begin{center}
\epsfigure{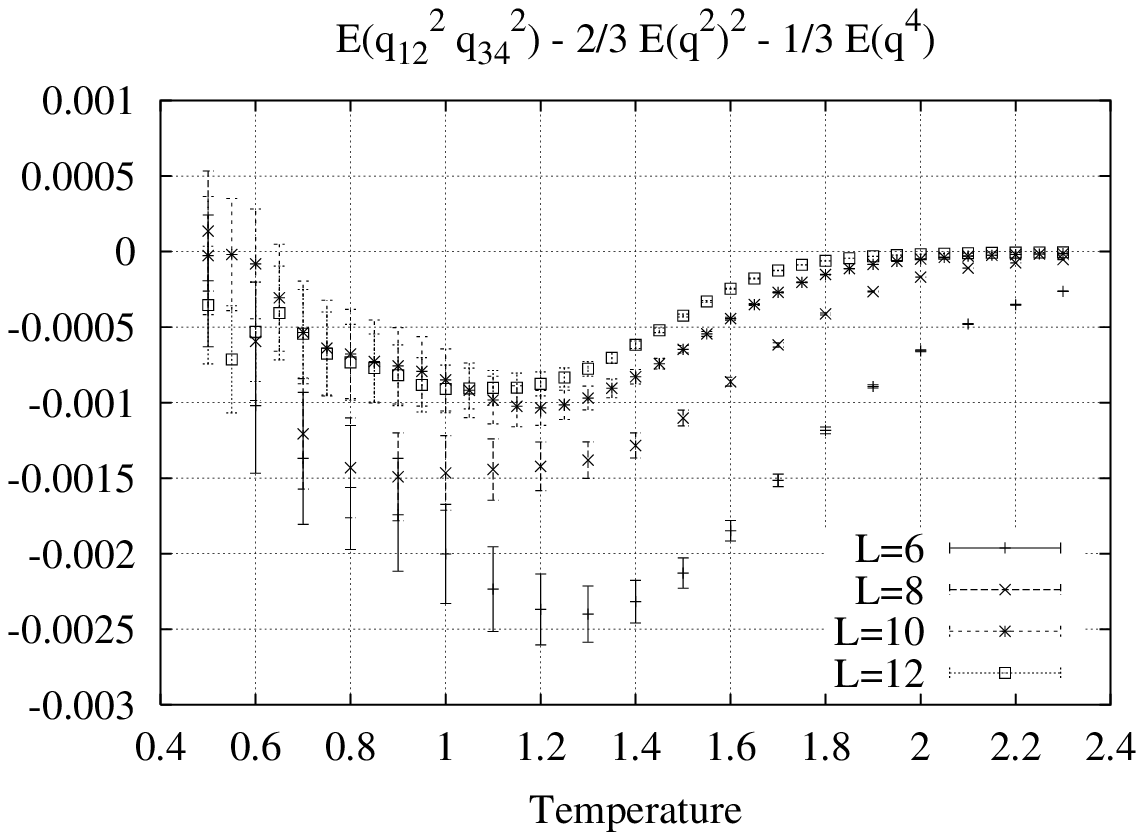}{3.5in}
\end{center}
\vspace{10pt}
\caption{The quantity $\ba{\lan q^2 \ran^{2}} - (\frac13 \ba{\lan q^4\ran}+
\frac23 \ba{\lan q^2\ran}^2)$ as function of the temperature for different values of $L$ in $D=3$.}
\label{G1}
\end{figure}

It useful to remark that if a symmetry is present, a system cannot be stochastically stable.  Indeed 
spin glasses may be stochastically stable only in the presence of a finite, non zero magnetic field 
which breaks the $\si \leftrightarrow -\si$ symmetry.  If a symmetry is present, stochastic 
stability may be valid only for those quantities which are invariant under the action of the 
symmetry group.  It is also remarkable that the union of two non-trivial uncoupled stochastically 
stable systems is {\sl not} stochastically stable.  Therefore a non-trivial stochastically stable 
system cannot be decomposed as the union of two or more parts whose interaction can be neglected.

In the general case stochastic stability implies  that
\be
P(q_{1},q_{2})\equiv\ba{P_{J}( q_{1}) P_{J}(q_{2})} 
 = \frac23 P(q_{1}) P(q_{2}) +
\frac13 P(q_{1}) \delta(q_{1}-q_{2}) \ . 
\ee
A particular case of the previous relation is the following one:
\be
\ba{\lan q^2 \ran^{2}}= \frac13 \ba{\lan q^4\ran}+
\frac23 \ba{\lan q^2\ran}^2\ . \label{CASE}
\ee

We have tested the previous relations in three dimensions as function of the temperature 
at different values of $L$ \cite{CINQUE}.  In fig.  \ref{G1} we plot the quantity $\ba{\lan q^2 
\ran^{2}} - (\frac13 \ba{\lan q^4\ran}+ \frac23 \ba{\lan q^2\ran}^2)$, which should be equal 
to zero. Indeed it is very small and its values decreases with $L$.
The two quantities in eq. (\ref{CASE}) in the low temperature region are a factor of $10^{3}$ bigger of their 
difference. I believe that there should be few doubts on the fact that stochastic 
stability is satisfied for three dimensional spin glasses.

\section*{Off-equilibrium dynamics}

The general problem that we face is to find what happens if the system is carried in a 
slightly off equilibrium situation. There are two ways in which this can be done.

    a) We rapidly cool the system starting from a random (high temperature) 
    configuration at time zero and we wait a time much longer than the microscopical one.  
    The system orders at distances smaller that a coherence distance $\xi(t)$ (which 
    eventually diverges when $t$ goes to infinity) but remains always disordered at 
    distances larger than $\xi(t)$.
    
    b) A second possibility consists in forcing the system in an off-equilibrium state  
    by gently {\it shaking} it. This can be done for example by adding a small time 
    dependent magnetic field, which should however strong enough force a large scale 
    rearrangement of the system \cite{TRE}. 

In the first case we have the phenomenon of ageing. This effects may be evidenziated if we  
define  a two time 
correlation function and two time relaxation function (we cool the system at time 0) 
\cite{cuku,frame}. 
The correlation function is defined to be
\be
C(t,t_w) \equiv \frac1N \sum_{i=1}^N \lan \si_i(t_w) \si_i(t_w+t)\ran\ ,
\ee
which is equal to the overlap $ q(t_{w},t_{w}+t)$ among a configuration at time $t_{w}$ and one at 
time $t_{w}+t$.
The relaxation function $S(t,t_w)$ is a just given by
\be
S(t,t_w)=\beta ^{-1}\lim_{\delta h \to 0 } {\delta m(t+t_{w}) \over \delta h}\ ,
\ee
where $\delta m$ is the variation of the magnetization when we add  a magnetic 
field $\delta h$ starting from time $t_{w}$. More generally we can introduce the time dependent 
Hamiltonian:
\be
H=H_{0}+\theta(t-t_{w}) \sum_{i} h_{i}\si_{i} \ .
\ee
The relaxation function is thus defined as:
\be
\beta S(t,t_w) \equiv \frac1N \sum_{i=1}^N  \lan {\partial \si_i(t_w+t)\over \partial h_{i}} \ran\ .
\ee

\begin{figure}[t] % fig 1
\begin{center}
\epsfigure{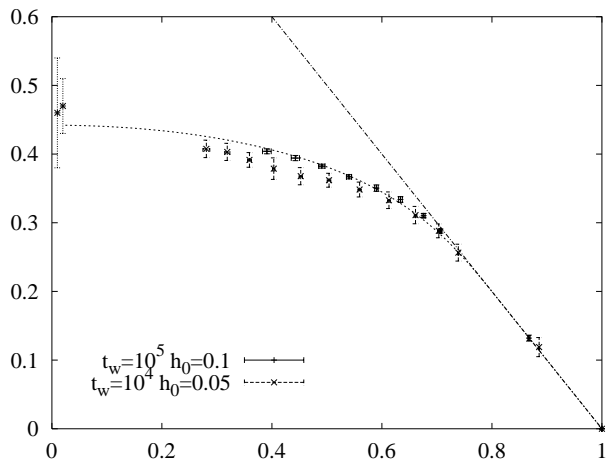}{3.5in}
\end{center}
\vspace{10pt}
\caption{Relaxation function versus correlation in the Edwards-Anderson (EA) model in $D=3$
     $T=0.7\simeq \frac34 T_c$ and theoretical predictions from eq. (\ref{FDR}) .}
\label{EA3}
\end{figure}

We can distinguish two situations:
\begin{itemize}
\item For $t<<t_w$ we stay in the {\sl quasi-equilibrium} regime \cite{FV}, $C(t,t_w) \simeq C_{\rm 
eq}(t)$, where $C_{\rm eq}(t)$ is the equilibrium correlation function; in this case $q_{EA} 
\equiv \lim_{t\to \infty} \lim_{t_w\to \infty} C(t,t_w)$.  

\item For $t=O(t_w)$ or larger we stay in 
the aging regime.  If simple aging holds $C(t,t_w) \propto {\cal C}(t/t_w)$.  
\end{itemize}

In the equilibrium regime, if we plot parametrically the relaxation function 
as function of the correlation, we find that
\be 
{ dS \over dC} =-1 \label{FDT}\ ,
\ee
which is a compact way of writing the fluctuation-dissipation theorem.

Generally speaking the fluctuation-dissipation theorem is not valid in the off-equilibrium regime.  
In this case one can use stochastic stability to derive a relation among statics properties and 
the form of the function $S(C)$ measured in off-equilibrium experiments \cite{cuku,frame,QUATTRO}:
\be
-{dS \over dC}=\int_{0}^{C} dq P(q)\equiv X(C) \ \label{FDR}.
\ee

The validity of these relation has been intensively checked in numerical experiments (see 
for example fig. (\ref{EA3}) from \cite{SG}).

In spin glasses the relaxation function has been experimentally measured many times in the aging 
regime, while the correlation function has not been measured: it is a much more difficult experiment 
in which one has to measure thermal fluctuations.  Fortunately enough measurements of both 
quantities for spin glasses are in now progress.  It would be extremely interesting to see if they 
agree with the theoretical predictions.

\end{document}